\documentclass{article}
\usepackage{spconf,amsmath,amssymb,amsfonts,graphicx,hyperref}
\usepackage{booktabs,multirow,array}
\usepackage[table]{xcolor}
\usepackage{microtype}
\usepackage{algorithm}
\usepackage{algpseudocode}
\usepackage{amsmath}
\usepackage{xcolor}

\newcommand{\bx}{\mathbf{x}}
\newcommand{\bq}{\mathbf{q}}
\newcommand{\bz}{\mathbf{z}}
\newcommand{\bU}{\mathbf{U}}
\newcommand{\bP}{\mathbf{P}}
\newcommand{\bR}{\mathbf{R}}
\newcommand{\bmu}{\boldsymbol{\mu}}
\newcommand{\AAm}{\textbf{AA}$\uparrow$}
\newcommand{\PDm}{\textbf{PD}$\downarrow$}
\newcommand{\best}[1]{\textbf{#1}}

\title{SPECTRA: Subspace-Preserving Embedding Calibration, Transport, and
Replay for Fully Few-Shot Class-Incremental Audio Classification}

\name{Giries Abu Ayoub\textsuperscript{1} \quad  Loay Mualem\textsuperscript{2,3} \quad  Simon Korman\textsuperscript{1}}
\address{\textsuperscript{1}Department of Computer Science, University of Haifa \\
  \textsuperscript{2}Institute for AI, University of Stuttgart \quad
    \textsuperscript{3}IMPRS-IS}

\begin{document}
\ninept
\maketitle

\begin{abstract}
Fully few-shot class-incremental audio classification (FFCAC) requires recognizing new sound classes from only a handful of labeled examples per
session, without forgetting previously learned classes and without any large
base dataset. Existing methods typically freeze a pre-trained
audio--language encoder and classify with point prototypes, but they suffer from significant performance degradation throughout the sessions due to generic feature representations. 

We propose \textbf{SPECTRA}, a framework built on a frozen encoder which adds three components.
(i) a lightweight trainable adapter that calibrates the generic
embeddings to the task; (ii) \emph{subspace feature replay}, an exemplar-free
anti-forgetting scheme that replays old classes by sampling from the low-rank
subspace of their stored features; and (iii) a transductive optimal-transport
refinement of prototypes at test time. Our central finding is that the
\emph{subspace structure} of the replay diminishes forgetting and
outperforms naive Gaussian replay of equal variance. On three FFCAC
benchmarks (NSynth-100, FSC-89, LS-100), SPECTRA improves average accuracy and
reduces forgetting over current state-of-the-art methods, and our ablations statistically validate each
component.
\end{abstract}


\section{Introduction}
\label{sec:intro}

Audio classification underpins applications from assisted driving~\cite{li2021spontaneous} and
medical monitoring~\cite{li2025end} to wildlife sensing~\cite{anuvind2023development}. In realistic
deployments, the set of classes is not fixed, since new categories appear over
time and must be learned from very few examples, without revisiting old
data. \emph{Few-shot class-incremental audio classification}
(FCAC) formalizes this, and its hardest variant,
\emph{fully} FCAC (FFCAC), removes the assumption of an abundant
base session: \emph{every} session, including the first, provides only $N$
classes with $K$ examples each. FFCAC therefore combines the two classic
difficulties of few-shot \emph{overfitting} and incremental
\emph{catastrophic forgetting}.

Pre-trained audio--language models (ALMs) such as
CLAP~\cite{elizalde2023clap} and PENGI~\cite{deshmukh2023pengi} provide
strong, transferable audio features. TAPE~\cite{gao2026tape} is the first
method to exploit an ALM for FFCAC and defines the current state of the
art. It \emph{freezes} the ALM encoder so representations stay stable across
sessions, and adds two lightweight components: a \emph{Task-Transform} that
maps features and prototypes into a task-adaptive metric space through a
closed-form orthogonalising transformation, and a \emph{Prototype Evolution}
step that refines each class prototype using low-entropy query samples.

\noindent\textbf{Related work.}
Classic few-shot class-incremental learning (FSCIL) methods decouple a frozen backbone from an evolving
classifier~\cite{tao2020few,zhang2021few}; in audio,
prototype-refinement~\cite{xie2023few} and capacity-growing methods address
FCAC and FFCAC, while SpurAudio~\cite{ayoub2026spuraudio} shows that few-shot audio classifiers readily latch onto spurious correlations, underscoring the need for task-specific representation calibration. A complementary line fights forgetting by \emph{replaying}
past data: rather than storing exemplars, generative and feature-replay
methods synthesise old-class samples~\cite{shin2017continual}, which is
attractive when raw data cannot be retained. On the inference side,
transductive methods such as PT-MAP~\cite{hu2021leveraging} exploit the joint statistics
of the query set through Sinkhorn optimal
transport~\cite{cuturi2013sinkhorn}. SPECTRA brings exemplar-free,
\emph{subspace-structured} feature replay and transductive transport to the
frozen-ALM pipeline of TAPE~\cite{gao2026tape}, where the tiny support sets make both
data-efficiency and forgetting acute.


\noindent\textbf{Motivation.}
Prototype-based representations provide an effective summary of each class for classification. However, in FFCAC the feature representation must continue to adapt as new classes arrive, while samples from previous sessions are no longer available. This raises a fundamental question: how can we continue adapting the representation without losing knowledge of earlier classes? Our key insight is that a compact representation of each class's feature geometry can be used to synthesize representative feature embeddings for replay, providing continual supervision for previous classes without storing raw audio.

\noindent\textbf{Contributions.}
Motivated by the need to continually adapt the feature representation while preserving knowledge of previously learned classes, we propose \textsc{SPECTRA}, a prototype-based framework that augments each class with a compact subspace representation to enable exemplar-free feature replay. Our main contributions are:
(i) A trainable residual-FFN \emph{adapter} that calibrates frozen ALM embeddings to the target task (Sec.~\ref{ssec:adapter}).
(ii) \emph{Subspace feature replay}, an exemplar-free anti-forgetting mechanism that models each class using a compact low-rank subspace and synthesizes representative feature embeddings for continual rehearsal (Sec.~\ref{ssec:replay}). We show that preserving subspace geometry is crucial, significantly outperforming naive replay strategies.
(iii) A \emph{transductive optimal-transport} refinement procedure that improves prototype estimation during inference (Sec.~\ref{ssec:ot}).
(iv) Extensive experiments on three FFCAC benchmarks demonstrating consistent improvements over TAPE in both accuracy and forgetting, supported by per-session analyses and statistical significance tests (Sec.~\ref{sec:exp}).

\begin{figure*}[t]
\centering
\includegraphics[width=0.75\textwidth]{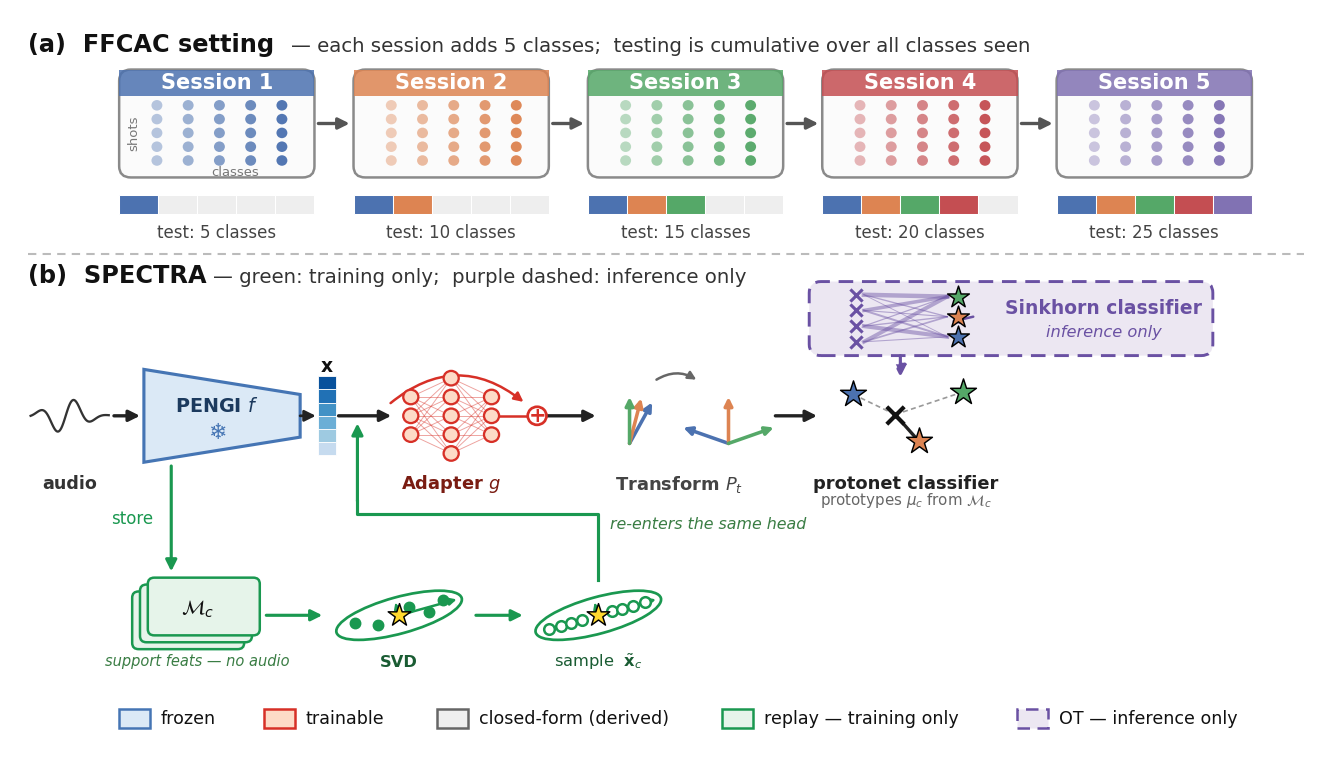}
\caption{\textbf{(a)} The FFCAC setting:
sessions arrive over time, each introducing $5$ new classes with $K{=}5$ shots
and \emph{no} base session; the model is evaluated on \emph{all} classes seen
so far (cumulative test set growing $5\!\to\!25$). \textbf{(b)} SPECTRA. An
audio clip is embedded by a \emph{frozen} audio--language encoder (PENGI $f$),
calibrated by a \emph{trainable} residual-MLP adapter $g$, and mapped by TAPE's
\emph{closed-form} orthogonalising transform $\bP_t$ before a prototypical
classifier. \emph{Green (training only):} each class's stored support features
$\mathcal{M}_c$ give a low-rank SVD subspace from which synthetic old-class
features $\tilde{\bx}_c$ are sampled and passed through the \emph{same} head, a
rehearsal that counters forgetting without storing audio. \emph{Purple, dashed
(inference only):} a transductive Sinkhorn step refines the prototypes from the
unlabelled query batch.}
\label{fig:illustration}
\end{figure*}

\section{Methodology}
\label{sec:method}
\subsection{Problem Setup}
\label{ssec:bg}

We consider the fully few-shot class-incremental audio classification (FFCAC)
setting. An audio-language model (ALM) encoder $f$ maps an audio clip to a
$d$-dimensional embedding $\bx\in\mathbb{R}^{d}$ ($d=1024$). Classes arrive
incrementally across sessions, where each new class is observed through only
$K$ support examples and samples from previous sessions are no longer
available. The goal is therefore to continually incorporate new classes while
preserving knowledge of previously learned ones under strict exemplar-free
constraints.

For each class $c$, we maintain the support feature set
$\mathcal{M}_c=\{\bx_i\}_{i=1}^{K}$ together with its prototype
$\bmu_c=\frac{1}{K}\sum_i\bx_i$. Following TAPE~\cite{gao2026tape}, we employ
its closed-form Task-Transform, which projects prototypes into an orthogonal
metric space prior to classification. This component is kept unchanged
throughout our method, allowing us to focus on improving representation
adaptation, continual rehearsal, and inference.

\subsection{Overview}
\label{ssec:overview}

Our framework consists of three complementary components. First, we introduce a
lightweight embedding adapter that calibrates frozen ALM features to the target
task without fine-tuning the backbone. Second, we preserve previous knowledge
through subspace feature replay, which models each class by a compact
low-dimensional subspace and synthesizes representative feature embeddings for
continual rehearsal. Finally, we refine class prototypes during inference using
transductive optimal transport over the unlabeled test batch. Figure~\ref{fig:illustration}
illustrates the complete pipeline.

\subsection{Embedding Adapter}
\label{ssec:adapter}

Frozen ALM embeddings provide strong generic representations but are not
optimized for the target taxonomy. Rather than fine-tuning the entire encoder,
which is prone to overfitting in the fully few-shot regime, we introduce a
lightweight residual adapter that calibrates the embedding space while
preserving the stability of the frozen backbone.

Specifically, we employ a transformer-style feed-forward residual block with
LayerScale,
$g(\bx)=\bx+\boldsymbol{\gamma}\odot \mathbf{W}_2\,\mathrm{GELU}\!\big(\mathbf{W}_1\,\mathrm{LN}(\bx)\big)$,
\label{eq:adapter}
where $\mathbf{W}_1\!\in\!\mathbb{R}^{rd\times d}$,
$\mathbf{W}_2\!\in\!\mathbb{R}^{d\times rd}$, and
$\boldsymbol{\gamma}$ is initialized with small values so that the adapter
starts close to the identity mapping while remaining expressive throughout
training.

The adapter is applied to both support and query embeddings before the
Task-Transform, i.e.,
$\bz=\bP_tg(\bx)$, ensuring that all features are represented in the same
calibrated embedding space. It is the only substantial trainable component of
our framework and is optimized end-to-end together with the classifier.

\subsection{Subspace Feature Replay}
\label{ssec:replay}

As the embedding adapter evolves to accommodate newly introduced classes,
previously learned classes become susceptible to forgetting. Since FFCAC
prohibits storing past audio, rehearsal must operate entirely in feature space.

Our key idea is to represent each class not only by its prototype, but also by
a compact low-dimensional subspace that captures its intrinsic feature
variation. This representation enables the synthesis of representative feature
embeddings, allowing previous classes to be rehearsed throughout continual
learning without retaining any raw audio.

Empirically, the support embeddings of a class occupy only a small region of
the feature space and are well approximated by a low-rank affine subspace. Let
$\tilde{\mathcal{M}}_c=[\bx_1-\bmu_c,\dots,\bx_K-\bmu_c]^\top$ with SVD
$\tilde{\mathcal{M}}_c=\bU\Sigma V^\top$. The top-$k$ singular vectors
$\bU_c=\bU_{:k}$ define the principal directions of the class, while
$\sigma_j=\Sigma_{jj}/\sqrt{K-1}$ captures the variation along each direction.

At session $t>0$, we generate $n$ pseudo embeddings for every previously seen
class by sampling within this subspace,
\begin{equation}
\tilde{\bx}_c=\bmu_c+\sum_{j=1}^{k} z_j\,(\bU_c)_j ,\qquad
z_j\sim\mathcal{N}\!\big(0,\sigma_j^2\big),
\label{eq:replay}
\end{equation}
The synthesized embeddings are processed by the current adapter and classifier
exactly as real samples. Consequently, replay regularizes the evolving
representation by encouraging previously learned classes to remain
discriminative as new classes are introduced.

\begin{figure}[t]
\centering
\includegraphics[width=\columnwidth]{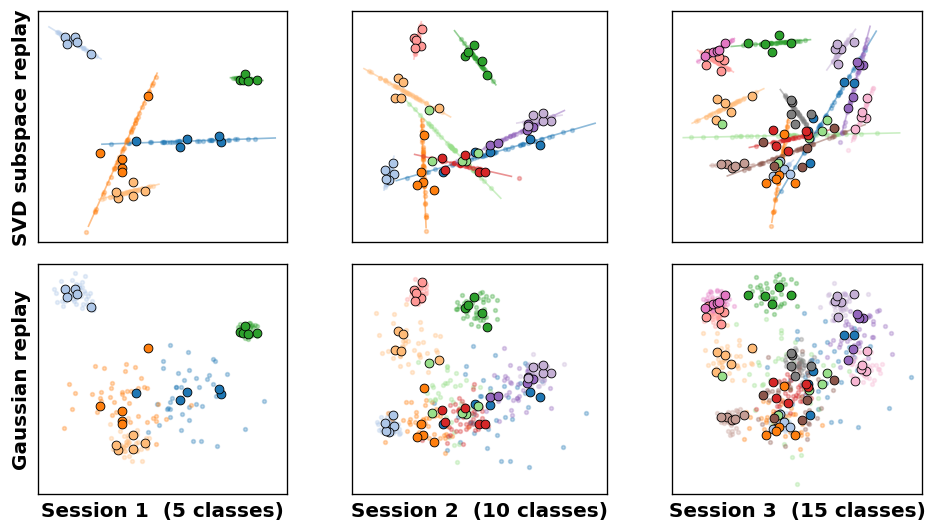}
\caption{Replay geometry across sessions. As more classes are added, low-rank subspace replay remains class-aligned, whereas Gaussian replay produces overlapping clouds.}
\label{fig:replay_comp2}
\end{figure}

The overall training objective combines the standard cross-entropy loss on the
current session with a replay loss over the synthesized embeddings,
\begin{equation}
\mathcal{L}=\mathcal{L}_{\text{new}}+\lambda\,
\mathcal{L}_{\text{replay}},\quad
\mathcal{L}_{\text{replay}}=\!\!\sum_{c<\,5t}\!\mathrm{CE}\big(\mathrm{head}(\tilde{\bx}_c),c\big).
\label{eq:replayloss}
\end{equation}
Unlike isotropic Gaussian replay, our approach preserves the principal
directions of variation within each class, producing substantially more
representative rehearsal samples. Section~\ref{ssec:abl} compares subspace
replay against Gaussian replay and replay-free training, and
Figure~\ref{fig:replay_comp2} visualizes the
difference between the two sampling strategies.

\subsection{Transductive Prototype Refinement}
\label{ssec:ot}

Few-shot prototypes are estimated from only $K$ support examples and therefore
provide imperfect estimates of the underlying class centers. Instead of
classifying each query independently, we refine the prototypes by exploiting
the collective structure of the unlabeled test batch through
entropy-regularized optimal transport.

Given the cosine-distance cost
$M_{jc}=1-\bq_j^\top\bmu_c$, Sinkhorn
iterations~\cite{cuturi2013sinkhorn} compute a soft transport plan
$\mathbf{R}=\operatorname*{Sinkhorn}_{\varepsilon}(M)$ under balanced row and
column constraints.

Each prototype is then updated using both its support embeddings and the
softly assigned query embeddings,
$\bmu_c \leftarrow
\frac{\sum_i \bx_i + \sum_j R_{jc}\,\bq_j}{K+\sum_j R_{jc}}$,
\label{eq:mstep}
We repeat this refinement for $T$ iterations before the final prediction.
By jointly considering all unlabeled queries, the refined prototypes provide a
more reliable estimate of the class centers than independent per-sample
classification, leading to improved recognition performance in the few-shot
setting.

\subsection{SPECTRA}
\label{ssec:full}

\noindent\textbf{Training:} We update only the adapter
$g$ and the learned prototypes and reference anchors; the encoder
$f$ is frozen and the transform $\bP_t$ is closed-form. The sole training objective is Eq.~\ref{eq:replayloss}: the new-class support features and the
synthetic replay features~(\ref{eq:replay}) are classified by cross-entropy.

\noindent\textbf{Inference (no gradients).} Only at test time do we apply the
transductive optimal transport step~(\ref{eq:mstep}): we push the query-set through the trained adapter and project it using the transform, run $T$ iterations to refine the prototypes from the unlabelled
queries, and classify. This step updates no parameters.
Algorithm~\ref{alg:spectra} summarises one session.

\section{Experiments}
\label{sec:exp}

\textbf{Setup.}
\label{ssec:setup}
We evaluate on NSynth-100~\cite{engel2017neural} (instrument notes), FSC-89~\cite{wang2021calls} (sound
events), and LS-100~\cite{panayotov2015librispeech} (speaker identity) under the standard FFCAC protocol: $5$ sessions, $5$ classes per
session, $K{=}5$ shots. We report \emph{average accuracy} (AA, mean over
sessions, $\uparrow$) and \emph{performance drop} (PD $=$ first $-$ last
session accuracy, $\downarrow$); all numbers are means over $50$ seeds. The
encoder is the frozen PENGI ALM and we keep TAPE's closed-form transform and
training otherwise untouched, reproducing its pipeline as our baseline.
Unless stated otherwise, the adapter expansion ratio is $r{=}3$, the replay subspace rank is $k{=}3$, the replay weight is $\lambda{=}1$, and OT uses $T{=}3$ prototype-refinement iterations.

\textbf{Main results.}
\label{ssec:main}
Tables~\ref{tab:nsynth_compact}--\ref{tab:ls100_compact} compare SPECTRA to
various FFCAC baselines. SPECTRA improves
average accuracy on \emph{all} datasets while
\emph{also} reducing forgetting (PD), confirming that task calibration plus
exemplar-free replay help on top of a strong frozen-ALM classifier.

\begin{table}[t]
\centering
\caption{Comparison of different methods on NSynth-100.}
\label{tab:nsynth_compact}
\small
\setlength{\tabcolsep}{1.8pt} 
\begin{tabular}{lccccccc} 
\toprule
Method & S0 & S1 & S2 & S3 & S4 & AA $\uparrow$ & PD $\downarrow$ \\
\midrule
$\text{iCaRL}_{\text{CVPR'20}}$~\cite{icarl}      & 71.70 & 53.51 & 53.66 & 49.07 & 49.48 & 55.48 & 40.08 \\
$\text{PODNET}_{\text{ECCV'20}}$~\cite{douillard2020podnet}     & 71.87 & 44.89 & 43.58 & 42.93 & 40.97 & 48.85 & 35.57 \\
$\text{DER}_{\text{CVPR'21}}$~\cite{der}        & 74.40 & 61.42 & 60.06 & 53.73 & 44.42 & 58.81 & 38.95 \\
$\text{CEC}_{\text{CVPR'21}}$~\cite{zhang2021few}        & 76.13 & 58.15 & 53.80 & 48.26 & 44.34 & 56.14 & 37.49 \\
$\text{FACT}_{\text{CVPR'22}}$~\cite{fact}      & 74.97 & 51.94 & 51.43 & 46.54 & 43.45 & 53.27 & 38.65 \\
$\text{PAN}_{\text{TMM'23}}$~\cite{pan}       & 76.71 & 58.38 & 53.92 & 48.44 & 44.48 & 56.39 & 37.72 \\
$\text{EDE}_{\text{Interspeech'24}}$~\cite{ede} & 76.16 & 70.18 & 63.46 & 59.16 & 58.02 & 65.40 & 18.14 \\
$\text{AISP}_{\text{TASLP'25}}$~\cite{aisp}     & 63.84 & 60.32 & 58.62 & 54.30 & 52.90 & 57.99 & 10.94 \\
\midrule
$\text{CLAP}_{\text{ICASSP'23}}$~\cite{elizalde2023clap}     & 43.80 & 22.14 & 16.15 & 12.69 & 12.31 & 21.42 & 31.49 \\
$\text{COOP}_{\text{IJCV'22}}$ ~\cite{coop}      & 78.35 & 44.45 & 30.99 & 23.23 & 17.91 & 38.98 & 60.45 \\
$\text{COCOOP}_{\text{CVPR'22}}$ ~\cite{cocoop}    & 69.60 & 39.79 & 27.24 & 21.53 & 16.67 & 34.96 & 52.94 \\
$\text{PALM}_{\text{EMNLP'24}}$~\cite{hanif2024palm}     & 94.09 & 57.34 & 41.43 & 34.08 & 31.32 & 51.65 & 62.77 \\
\midrule
$\text{TAPE}_{\text{CVPR'26}}$~\cite{gao2026tape}      & 96.64 & 94.53 & 93.44 & 91.72 & 91.05 & 93.48 & 5.58 \\
\rowcolor{green!7}
$\textbf{SPECTRA}_{\textbf{(ours)}}$ & \textbf{98.14} & \textbf{97.27} & \textbf{96.77} & \textbf{95.50} & \textbf{94.92} & \textbf{96.52} & \textbf{3.23} \\
\bottomrule
\end{tabular}
\end{table}

\begin{table}[t]
\centering
\caption{Comparison of different methods on FSC-89.}
\label{tab:fsc89_compact}
\small
\setlength{\tabcolsep}{1.8pt} 
\begin{tabular}{lccccccc}
\toprule
Method & S0 & S1 & S2 & S3 & S4 & AA $\uparrow$ & PD $\downarrow$ \\
\midrule
$\text{PAN}_{\text{TMM'23}}$~\cite{pan}         & 41.48 & 23.72 & 18.08 & 15.27 & 12.25 & 22.16 & 37.72 \\
$\text{EDE}_{\text{Interspeech'24}}$~\cite{ede} & 53.25 & 37.65 & 35.59 & 32.69 & 27.60 & 37.36 & 30.05 \\
$\text{AISP}_{\text{TASLP'25}}$~\cite{aisp}      & 62.66 & 48.92 & 43.81 & 37.73 & 33.43 & 45.31 & 29.23 \\
\midrule
$\text{CLAP}_{\text{ICASSP'23}}$~\cite{elizalde2023clap}     & 49.64 & 37.73 & 28.55 & 25.42 & 23.70 & 33.01 & 25.94 \\
$\text{COOP}_{\text{IJCV'22}}$~\cite{coop}       & 63.32 & 37.75 & 27.50 & 22.07 & 18.64 & 33.86 & 44.69 \\
$\text{COCOOP}_{\text{CVPR'22}}$~\cite{cocoop}     & 64.65 & 37.62 & 26.63 & 21.22 & 16.85 & 33.39 & 47.80 \\
$\text{PALM}_{\text{EMNLP'24}}$~\cite{hanif2024palm}      & 74.41 & 39.16 & 25.30 & 19.91 & 17.47 & 35.25 & 56.94 \\
\midrule
$\text{TAPE}_{\text{CVPR'26}}$~\cite{gao2026tape}      & 81.24 & 72.65 & 67.07 & 63.93 & 61.29 & 69.24 & 19.96 \\
\rowcolor{green!7}
$\textbf{SPECTRA}_{\textbf{(ours)}}$ & \textbf{82.24} & \textbf{74.41} & \textbf{69.25} & \textbf{66.21} & \textbf{63.48} & \textbf{71.12} & \textbf{18.77} \\
\bottomrule
\end{tabular}
\end{table}

\begin{table*}[t]
\centering
\caption{Per-session accuracy (\%) under harder protocols: $10$-way $\times\,10$-session
on NSynth-100 and LS-100, and $10$-way $\times\,8$-session on FSC-89
(FSC-89 has only $89$ classes, capping it at $8$ sessions); $30$ seeds.}
\label{tab:tenway}
\small
\setlength{\tabcolsep}{4pt}
\begin{tabular}{l cccccccccc cc}
\toprule
Method & S0 & S1 & S2 & S3 & S4 & S5 & S6 & S7 & S8 & S9 & \AAm & \PDm \\
\midrule
\multicolumn{13}{l}{\textbf{NSynth-100} \;($10$-way $\times\,10$-session)}\\
TAPE & 95.6 & 92.2 & 90.8 & 89.6 & 88.4 & 86.5 & 85.1 & 84.0 & 82.9 & 81.5 & 87.66 & 14.11 \\
\rowcolor{green!7}
\textbf{SPECTRA} & \best{97.6} & \best{95.6} & \best{94.3} & \best{93.4} & \best{92.7} & \best{91.5} & \best{90.8} & \best{90.1} & \best{89.6} & \best{88.9} & \best{92.45} & \best{8.77} \\
\midrule
\multicolumn{13}{l}{\textbf{LS-100} \;($10$-way $\times\,10$-session)}\\
TAPE & 90.1 & 85.4 & 80.7 & 77.9 & 75.2 & 73.4 & 70.7 & 69.0 & 67.2 & 65.7 & 75.53 & 24.43 \\
\rowcolor{green!7}
\textbf{SPECTRA} & \best{95.1} & \best{91.2} & \best{88.1} & \best{85.7} & \best{83.7} & \best{82.3} & \best{80.4} & \best{79.7} & \best{78.6} & \best{77.8} & \best{84.26} & \best{17.31} \\
\midrule
\multicolumn{13}{l}{\textbf{FSC-89} \;($10$-way $\times\,8$-session)}\\
TAPE & 71.7 & 63.2 & 58.1 & 55.0 & 51.9 & 49.6 & 47.1 & 45.0 & -- & -- & 55.22 & 26.66 \\
\rowcolor{green!7}
\textbf{SPECTRA} & \best{74.1} & \best{65.9} & \best{60.7} & \best{57.4} & \best{54.5} & \best{52.5} & \best{50.3} & \best{48.7} & -- & -- & \best{58.02} & \best{25.44} \\
\bottomrule
\end{tabular}
\end{table*}

\begin{table}[t]
\centering
\caption{Comparison of different methods on LS-100.}
\label{tab:ls100_compact}
\small
\setlength{\tabcolsep}{1.8pt} 
\begin{tabular}{lccccccc}
\toprule
Method & S0 & S1 & S2 & S3 & S4 & AA $\uparrow$ & PD $\downarrow$ \\
\midrule
$\text{PAN}_{\text{TMM'23}}$~\cite{pan}         & 85.70 & 52.20 & 39.17 & 32.95 & 29.88 & 47.98 & 37.72 \\
$\text{EDE}_{\text{Interspeech'24}}$~\cite{ede} & 91.90 & 70.23 & 54.21 & 46.97 & 45.94 & 61.85 & 45.96 \\
$\text{AISP}_{\text{TASLP'25}}$~\cite{aisp}      & 91.20 & 70.26 & 53.71 & 47.08 & 45.15 & 61.48 & 46.05 \\
\midrule
$\text{CLAP}_{\text{ICASSP'23}}$~\cite{elizalde2023clap}     & 18.96 & 9.06  & 5.91  & 4.75  & 3.37  & 8.41  & 15.59 \\
$\text{COOP}_{\text{IJCV'22}}$~\cite{coop}       & 49.14 & 27.64 & 18.48 & 14.48 & 11.01 & 24.15 & 38.14 \\
$\text{COCOOP}_{\text{CVPR'22}}$~\cite{cocoop}     & 46.90 & 24.32 & 15.75 & 10.86 & 7.84  & 21.14 & 39.06 \\
$\text{PALM}_{\text{EMNLP'24}}$~\cite{hanif2024palm}      & 87.73 & 44.96 & 29.39 & 21.44 & 18.97 & 40.50 & 68.76 \\
\midrule
$\text{TAPE}_{\text{CVPR'26}}$~\cite{gao2026tape}      & 92.34 & 87.41 & 84.64 & 82.24 & 80.84 & 85.49 & 11.50 \\
\rowcolor{green!7}
$\textbf{SPECTRA}_{\textbf{(ours)}}$ & \textbf{97.22} & \textbf{94.81} & \textbf{92.32} & \textbf{89.99} & \textbf{88.55} & \textbf{92.58} & \textbf{8.67} \\
\bottomrule
\end{tabular}
\end{table}

\textbf{Per-session analysis.}
\label{ssec:persess}
The per-session columns of
Tables~\ref{tab:nsynth_compact}--\ref{tab:ls100_compact} show \emph{where} the
gains arise: SPECTRA leads at every session, and on NSynth-100 the margin
\emph{grows} over sessions (S0 $+1.5$, S4 $+3.9$ over TAPE). The effect is largest on LS-100 ($+7.1$ AA).


\textbf{Scaling to a harder protocol.}
\label{ssec:tenway}
To test generality beyond the standard $5$-way protocol, we run a harder
$10$-way $\times\,10$-session setting on all datasets (all dataset classes over
$10$ incremental steps). Table~\ref{tab:tenway} shows SPECTRA still
beats TAPE by a wide margin.

\begin{algorithm}[t]
\caption{SPECTRA: training and inference at session $t$}
\label{alg:spectra}
\small
\begin{algorithmic}[1]
\Require frozen ALM $f$; adapter $g$; anchors $\mathbf{A}$; memories $\{\mathcal{M}_c\}$
\Ensure labels $\hat{y}$ for queries of all classes seen so far
\Statex \textbf{Training} \Comment{only $g$ and the head get gradients}
\State $\mathcal{M}_c \gets \{f(a_i)\}_{i=1}^{K}$, \; $\bmu_c \gets \tfrac{1}{K}\sum_i \bx_i$ \Comment{new classes}
\For{each old class $c$}
  \State $\bU_c, \Sigma \gets \mathrm{SVD}(\mathcal{M}_c - \bmu_c)$, \; $\sigma_j \gets \Sigma_{jj}/\sqrt{K\!-\!1}$
  \State $\tilde{\bx}_c \gets \bmu_c + \sum_{j\le k} z_j \sigma_j (\bU_c)_j$, \; $z_j \!\sim\! \mathcal{N}(0,1)$ \Comment{Eq.~\ref{eq:replay}}
\EndFor
\State $\mathcal{L} \gets \mathcal{L}_{\mathrm{new}}(\bx) + \lambda \mathcal{L}_{\mathrm{replay}}(\tilde{\bx})$ \Comment{Eq.~\ref{eq:replayloss}}
\State $g,\,\mathrm{head} \gets \mathrm{SGD}(\mathcal{L})$
\State $\bP_t \gets \mathrm{pinv}(C)\,\mathbf{A}$ \Comment{closed form}
\Statex \textbf{Inference} \Comment{no gradients}
\State $\bq_j \gets \ell_2(\bP_t\,g(f(a_j)))$, \; $\bmu_c \gets \ell_2(\bP_t\,g(\bmu_c))$
\For{$T_{\mathrm{ot}}$ iterations}
  \State $M_{jc} \gets 1 - \bq_j^{\top}\bmu_c$, \; $\bR \gets \mathrm{Sinkhorn}_{\varepsilon}(M)$
  \State $\bmu_c \gets \ell_2\!\big(\bmu_c + \sum_j R_{jc}\bq_j\big)$ \Comment{Eq.~\ref{eq:mstep}}
\EndFor
\State \Return $\hat{y}_j = \arg\max_c \cos(\bq_j, \bmu_c)$
\end{algorithmic}
\end{algorithm}

\begin{table}[t]
\centering
\caption{Component ablation (AA / PD, 50 seeds).}
\label{tab:abl}
\setlength{\tabcolsep}{4pt}
\begin{tabular}{lcccc}
\toprule
\multirow{2}{*}{Configuration}
 & \multicolumn{2}{c}{NSynth-100} & \multicolumn{2}{c}{FSC-89}\\
\cmidrule(lr){2-3}\cmidrule(lr){4-5}
 & \AAm & \PDm & \AAm & \PDm \\
\midrule
TAPE (baseline)        & 93.5 & 5.6 & 69.2 & 20.0 \\
\;+ Adapter            & 95.5 & 4.3 & 70.0 & 19.0 \\
\;+ Transport (OT)     & 95.8 & 4.7 & 70.5 & 19.9 \\
\;\;\;w/ Gaussian replay & 95.8 & 4.6 & 70.8 & 19.3 \\
\rowcolor{green!7}
\;\;\;w/ subspace replay & \best{96.5} & \best{3.2}
                         & \best{71.1} & \best{18.8} \\
\bottomrule
\end{tabular}
\end{table}

\textbf{Ablations.}
\label{ssec:abl}
\noindent\textit{(i) Component contribution.}
Table~\ref{tab:abl} decomposes SPECTRA. Removing the adapter gives up the largest single drop in AA and
already lowers PD, so calibrating the frozen embedding is the main
driver; optimal transport then adds accuracy at a small PD cost (discussed
below); subspace replay yields the best AA \emph{and} the lowest PD.

\noindent\textit{(ii) Does the replay structure matter?}
We compare our low-rank replay~(\ref{eq:replay}) against (i) \emph{no replay}
and (ii) an \emph{isotropic Gaussian} rehearsal of matched total variance,
$\tilde{\bx}_c=\bmu_c+\boldsymbol\epsilon$,
$\boldsymbol\epsilon\!\sim\!\mathcal{N}(0,\tau^2\mathbf{I}_d)$,
$\tau^2=\tfrac1d\sum_j\sigma_j^2$. Subspace replay beats Gaussian replay on
2 datasets and both metrics. On NSynth-100 Gaussian replay barely improves over
no replay, isolating the \emph{anisotropic, low-rank structure}---not the act
of replaying---as the source of the anti-forgetting effect.

\noindent\textit{(iii) Adapter capacity.}
Sweeping the expansion $r\!\in\!\{2,3,4,8\}$ moves AA by less than the seed
standard deviation on both datasets; we use $r{=}3$, the smallest value on this
plateau.


\noindent\textit{(iv) Transport assignment.}
Optimal transport is the one component with a trade-off: across the chain it
raises AA ($+0.3$/$+0.5$) but increases PD ($+0.4$/$+0.9$). Subspace replay
more than compensates, yielding the best PD overall; we therefore retain OT
for its accuracy gain while replay supplies the anti-forgetting.

\section{Conclusion}
\label{sec:conc}
We presented SPECTRA for fully few-shot class-incremental audio
classification: a trainable adapter that calibrates frozen audio--language
embeddings, exemplar-free subspace feature replay that fights forgetting by
sampling old classes from their low-rank feature subspaces, and a
transductive optimal-transport refinement. On three FFCAC benchmarks SPECTRA
improves average accuracy and reduces forgetting over TAPE the current state of the
art, and our ablations statistically establish that the \emph{subspace
structure} of the replay---not replay alone---drives the gain. Future work
includes coupling the transport plan with the replay subspaces to further
stabilise noisier datasets.


\bibliographystyle{IEEEbib}
\bibliography{spectra}

\end{document}